\documentclass[finalold]{agujournal2019}
\usepackage{url} 
\usepackage[inline]{trackchanges} 
\usepackage{soul}

\usepackage{amsmath}
\usepackage{amssymb}

\newcommand{\vv}[1]{\mathbf{#1}}
\newcommand{\ii}[1]{\mathrm{#1}}

\journalname{JGR: Space Physics}

\begin{document}

%
%

\title{Resonant whistler-electron interactions: MMS observations vs. test-particle simulation}

%
%




\authors{E. Behar\affil{1}, F. Sahraoui\affil{1} and L. Ber\v{c}i\v{c}\affil{2,3}}


\affiliation{1}{Laboratoire de Physique des Plasmas, CNRS - \'Ecole Polytechnique - CNRS - Sorbonne Université - Unversit\'e Paris Saclay- Observatoire de Paris-Meudon, F-91128 Palaiseau Cedex, France}
\affiliation{2}{LESIA, Observatoire de Paris, PSL Research University, CNRS, UPMC Université Paris 6, Université Paris-Diderot, Meudon, France}
\affiliation{3}{Physics and Astronomy Department, University of Florence, Sesto Fiorentino, Italy}




\correspondingauthor{Etienne Behar}{etienne.behar@lpp.polytechnique.com}




\begin{keypoints}
\item Characteristic double-branch signatures in the electron Velocity Distribution Function (VDF) are observed simultaneously with a whistler wave.
\item The wave, applied to test-particles, produces signatures in the VDF through Landau and cyclotron resonances.
\item This resonant wave-particle interaction cannot be diagnosed in the Magnetospheric MultiScale (MMS) observations through the dissipative $\mathbf{E}\cdot\mathbf{J}$ term.
\end{keypoints}

%
%

%
%


\begin{abstract}
    Simultaneous observation of characteristic 3-dimensional (3D) signatures in the electron velocity distribution function (VDF) and intense quasi-monochromatic waves by the Magnetospheric Multiscale (MMS) spacecraft in the terrestrial magnetosheath are investigated. The intense wave packets are characterised and modeled analytically as quasi-parallel circularly-polarized whistler waves and applied to a test-particle simulation in view of gaining insight into the signature of the wave-particle resonances in velocity space. Both the Landau and the cyclotron resonances were evidenced in the test-particle simulations. The location and general shape of the test-particle signatures do account for the observations, but the finer details, such as the symmetry of the observed signatures are not matched, indicating either the limits of the test-particle approach, or a more fundamental physical mechanism not yet grasped. Finally, it is shown that the energisation of the electrons in this precise resonance case cannot be diagnosed using the moments of the distribution function, as done with the classical ${\bf E}.{\bf J}$ ``dissipation'' estimate.
\end{abstract}


\section{Introduction}

    Resonant wave-particle interactions are one of the few mechanisms in collisionless plasmas that enable a net transfer of energy from oscillating electromagnetic fields to moving charged particles. They play a fundamental role in various regions of the near-Earth plasma environment, such as the bowshock, the radiation belts, the polar cusp or the magneto-tail \cite{Mazelle00,Grison05, thorne2010grl,fujimoto2011ssr,krasnoselskikh2013ssr}. In (fully developed) plasma turbulence, wave-particle interactions are also thought to play a leading role in dissipating energy as the turbulent cascade proceeds from large (fluid) to small (kinetic) scales \cite{bruno2013lrsp, Sahraoui20}. In the solar wind (and to some extent the magnetosheath), the most debated dissipation processes are the Landau damping \cite{Landau46,Howes08,Schekochihin09,Gary09,Sahraoui10,Podesta10,Sulem15,Kobayashi17}, cyclotron damping \cite{Leamon98,Kasper08,Cranmer14,he2015apj} and stochastic heating \cite{chandran10}, which all would involve different spatial or temporal scales. Often magnetic reconnection is also evoked as a potential dissipation process in localized current sheets that self-consistently form in turbulence plasmas \cite{Matthaeus84,Retino07,Sundkvist07,Chasapis15}. However, even within such localized coherent structures, Landau damping is shown to be very effective in numerical simulations of collisionless magnetic reconnection \cite{tenbarge13,Loureiro13,numata15}.  \\

    Despite their role in energy dissipation, a {\it direct} diagnosis of wave-particle resonances in numerical simulations and in-situ data remains elusive. The difficulty to approach these processes stems from the need to measure {\it simultaneously} the 7D VDF (3 spatial dimensions, 3 velocity dimensions, and time) with high temporal and velocity space resolutions to access the kinetic scales, and the 4D structure of the electric and magnetic fields. While the latter could have been achieved at the magnetohydrodynamic (MHD) and ion scales using the Cluster data and appropriate data analysis techniques \cite{Sahraoui06,Narita10,Sahraoui10}, the former became possible only in recent years thanks to the MMS mission \cite{burch2016ssr}. MMS indeed provides us with the highest ever achieved resolution of the particle VDFs, both in time and velocity space \cite{pollock2016ssr}. Furthermore, thanks to its small inter-spacecraft separations ($\sim 10$ km) MMS allows us to probe in 3D kinetic spatial scales of the fluctuations fields. On the other hand, the increasing computer capabilities allows achieving Vlasov simulations with high-enough phase space resolution to unravel the complex nature of the kinetic dissipation in turbulent collisionless plasmas \cite{Cerri18}. The present article is part of the ongoing efforts in this direction. \\


    A few observational approaches on signatures of such mechanisms were previously proposed, and here we focus on studies manipulating 3-dimensional VDFs. \citeA{kitamura2018science} reported the unambiguous observation of wave-ion resonances leading to particle acceleration, associated with an ion cyclotron wave, in the Earth magnetosphere, displaying clear agyrotropic signatures (phase bunching) and their time evolution. In the work of \citeA{gershman2017nc}, at the magnetopause, the energy exchange between electrons and kinetic Alfv\'en waves was studied in terms of the dissipative $\mathbf{E}.\mathbf{J}$ term, and trapped electrons were found in the wave minima. In the radiation belts, \citeA{min2014jgr} proposed empirical indications for Landau resonance signatures in the electron VDFs, as local minima in their velocity derivatives close to the parallel resonant velocity, in the presence of Chorus waves. \citeA{he2015apj} and \citeA{marsch2011ag} show indications in the solar wind of resonances in the proton VDFs, composed of a diffused, anisotropic core and secondary beam, which were linked to kinetic waves. In the Earth magnetosheath and using a field-particle correlation technique, \citeA{chen2019nc} presented structures in the fluctuating electron VDF close to the electron therrmal speed, which the authors linked to electron Landau damping. \\

    The method followed in the present study is similar to the approach of \citeA{kitamura2018science} and \citeA{gershman2017nc}, in that we first identify a neat electromagnetic wave, as intense and monochromatic as possible, study its potential effect on particle Velocity Distribution Functions (VDF), and then compare the expected signatures with the VDF observed outside and inside the region where the wave is observed. This approach enables an unequivocal, 3-dimensional comparison of resonant signatures in the VDF between observations and the simulation.

    The wave studied in this work is a high-frequency quasi-parallel whistler mode, ubiquitous in both magnetospheric and the solar wind plasmas \cite{tao2012grl, lacombe2014apj, stansby2016apj}. Electrons are therefore the species of interest, and the frequencies of both the wave and the electron motion (gyration) are much higher than the fastest particle instruments operating in space, though in the reach of wave sensors. Anisotropies in the electron distribution functions are fundamental for the generation of whistler waves, as shown in various contexts such as the solar wind \cite{tong2019apj}, magnetic reconnection regions \cite{Huang16, yoo2018grl, yoo2019pp}, or mirror mode magnetic holes or other coherent structures in the magnetosheath \cite{Huang17, ahmadi2018jgr}. \citeA{verscharen2019apj}, \citeA{vocks2005apj} and \citeA{seough2015apj} have explored the theoretical link between whistler waves and solar wind electrons, in either the formation or the scattering of strahl and halo electrons. In their numerical approach, \citeA{hsieh2017jgr} show how oblique whistler mode chorus in the magnetosphere can lead to electron acceleration up to a few MeV, via Landau, cyclotron, and higher order resonances.

    To go beyond the 1-dimensional description of wave-particle resonances, in which a resonance is reduced to its associated (scalar) parallel speed, we explore the possibilities offered by a test-particle approach for a more comprehensive description of the mechanism and a direct comparison with the observations. This approach also presents the great advantage of isolating the effect of the wave on the particles, with no feedback allowed. A succinct view on particle energisation is also proposed, in order to appreciate whether the energy gained by the resonant particles can be quantified in observations.

\section{Particle data analysis}\label{sec:partAnalysis}

\begin{figure}
        \centering
        \includegraphics[width=\textwidth]{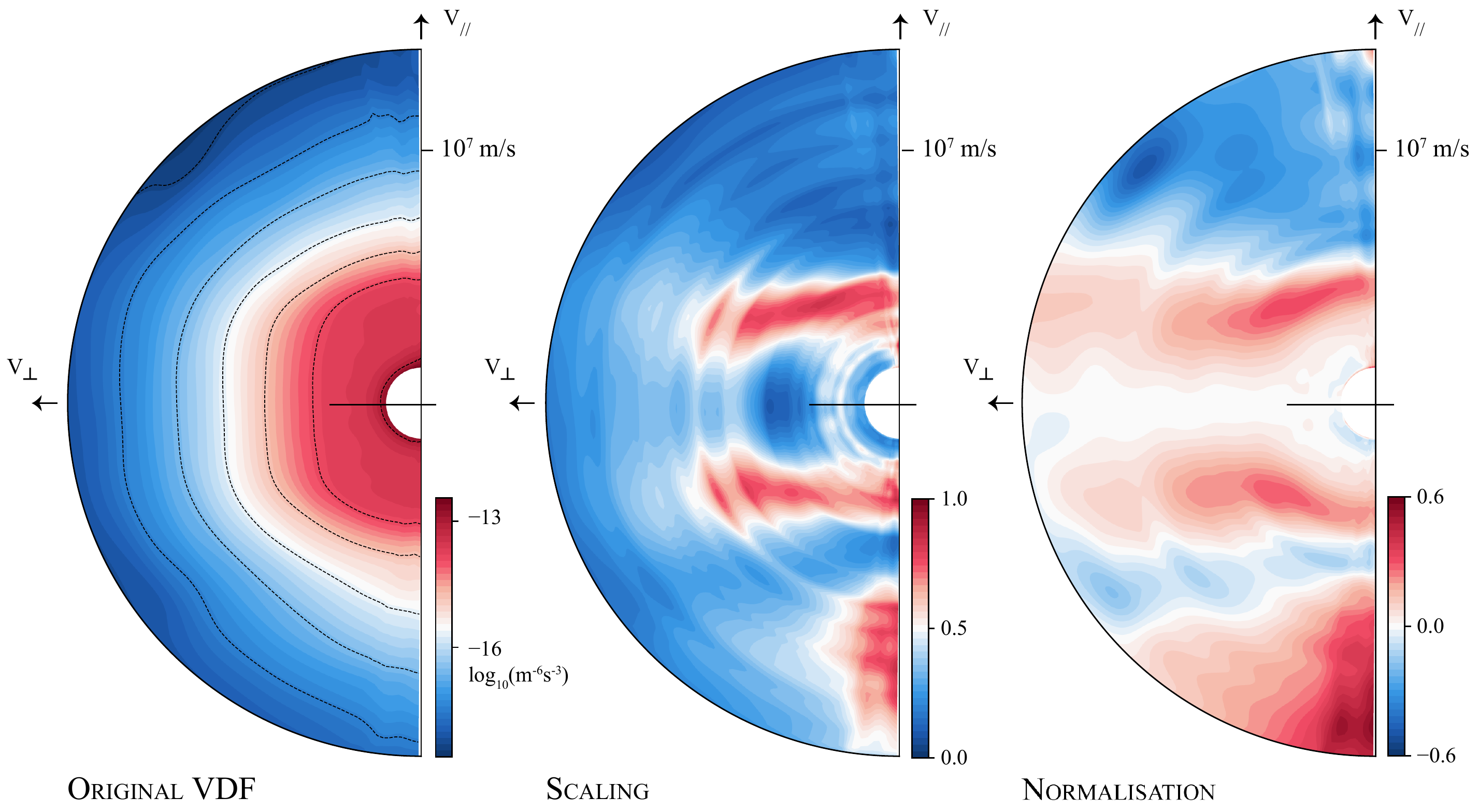}
        \caption{An example of an original VDF to the left, integrated over 300 ms, its scaled view in the middle, and its normalized view to the right.}
        \label{fig:VDFNorms}
    \end{figure}

    The particle data used in this study were recorded by the Fast Plasma Investigation (FPI) of the Magnetospheric Multiscale (MMS) mission \cite{burch2016ssr, pollock2016ssr}. We work in a reference frame in which the average plasma flow velocity $\vv{u}(t) = \frac{m_\ii{e}\vv{u}_\ii{e}(t) + m_\ii{i}\vv{u}_\ii{i}(t)}{m_\ii{e} + m_\ii{i}}$ is zero, and the {\it local} magnetic field ${\bf B}_0$ -- averaged over each 30 ms FPI measurement -- is aligned with the $z$-direction. We define a spherical grid-of-interest in this reference frame, of arbitrary extent and resolution. For each FPI measurement of the VDF, this grid-of-interest is rotated and shifted to the instrument reference frame (cf. \ref{app:binningInterpolation}) using the measured background magnetic field ${\bf B}_0$ and flow velocity, as well as the probe motion. The VDF is then interpolated at each node of the transformed grid, using a tricubic interpolation scheme, documented and tested in \ref{app:interpolation}. The use of a spherical grid allows us to represent an averaged VDF in the $(v_\perp,v_\parallel)$-plane without the need of a binning process, which is a source of systematic artifacts when working with multidimensional data. These aspects are illustrated in \ref{app:binningInterpolation}. In Figure \ref{fig:VDFNorms}, the left-hand polar plot shows the result for 10 time-averaged VDFs, with the parallel velocity given along the vertical axis. A regular spherical grid-of-interest of 200x200x200 nodes was used with a maximum extent of $1.5 \cdot 10^7$ m/s. To further ease the reading of the plots, a filled-contour representation is used.

    Our goal is to study how the shape of the VDF may be affected by the presence of a wave. We wish to go further than the reduced description of the VDF given by its first order moments, namely its number density, its number flux density ($\sim$ bulk velocity) and its momentum flux density tensor ($\sim$ temperature in the thermal equilibrium case). For this purpose, we need a process that enhances potential patterns which might be ``hidden'' by the order zero distribution, or \emph{background} distribution (not necessarily Gaussian/Maxwellian). In the example used in Figure \ref{fig:VDFNorms}, the VDF stretches over more than 6 orders of magnitude in the covered velocity space. The most obvious, order-0 shape found in the raw, original VDF is a centered, somewhat isotropic peak. A closer look may reveal obvious departure from the isotropy, with noticeably straight isocontours for parallel velocities around 0 m/s. We want a process which highlights these characteristics. 

    In a numerical context, such a background, or equilibrium distribution $f_0$ is usually subtracted from $f(t)$, with the similar goal of enhancing higher order features (in such a case, the departure from $f_0$). It is often defined as the initial VDF $f(t=0)$, and sometimes as the time averaged distribution $<f(t)>$. But the resulting \emph{fluctuating} distribution $\delta f(t) = f(t) - f_0$ might not be a valid concept when dealing with observations, as the \emph{background}, order-0 distribution itself is generally varying -- slightly or significantly -- during the time interval of interest. In other words, as the plasma flows and the spacecraft moves, we never probe plasmas with the same parameters, which may be the case in a controlled simulation box. Using such a $f_0$ with observations results in significant patterns in velocity space, which should be avoided for studying instantaneous, fine details of the VDF.

    With this motivation, we propose two different treatments of the VDF which do not rely on any other information than the instantaneous distribution itself. These two treatments provide two complementary views of the VDF, with different advantages and drawbacks discussed in the following sections. The first treatment is a \emph{scaling}, during which we consider each energy level of the spherical grid-of-interest separately. For each of these spherical shells, the minimum value of the interpolated VDF is set to 0 and the maximum value set to 1. Values are then averaged over the gyro-angle (angle around the background magnetic field). This scaling is closely related to classical pitch-angle distributions, in which the VDF is shown for a few selected energy ranges, with the color-plot dynamics ranging from the lowest to the highest distribution value of each energy range (see also \ref{app:interpolation}). In the proposed scaling, we virtually display 200 concentric, time-averaged pitch-angle distributions. In the next section, we will use a \emph{scaled} pitch-angle distribution, by selecting only one limited energy range of the scaled VDF and plotting it over time (Figure \ref{fig:spectroVDF}).

    In the second treatment, each VDF values is \emph{normalized} to a reference value, which we choose to be the VDF value at the same energy for $v_\parallel=0$ (i.e., a pitch-angle of $90^\circ$). All these central (equatorial) values are therefore set to one, appearing in white tones, while higher values appear as red tones and lower values as blue. A decimal logarithm is applied to the normalized values when plotted.

    These two treatments do not have a physical motivation, they are only arbitrarily chosen to highlight VDF features.

    In Figure \ref{fig:VDFNorms}, the effect of the scaling on the original distribution is remarkable, with two parallel branches of values larger than 0.5 (i.e. red tones) stretching at constant parallel velocities. Properly speaking, these two structures are thick disks in the 3-dimensional velocity space. Another structure is found along the anti-parallel direction, a strahl-like structure greatly highlighted in comparison with the original VDF. The circular features of constant speed are unavoidable artifacts of this scaling, in which all energy levels are treated regardless of the others: the continuity of the VDF across energy is lost, and some energy shells are scaled differently compared to the neighbouring shells, resulting in circular visual artifacts. The normalization we introduced here-above was applied on the original VDF, in order to conserve this continuity across energy, for a complementary representation given in the right-most plot of Figure \ref{fig:VDFNorms}. The circular artifacts vanished from the two branches -- or disks -- of higher density, which are now reaching to even higher perpendicular velocities. 
    In this view, the strahl-like component is perceived as broader than in the scaled VDF. The normalization puts different weights on details compared to the scaling, and though being much less sharp, it appears in this case to be more sensitive to details at high velocities.

    We now have a comprehensive and constraining way of examining an instantaneous VDF, with two different views of it. The 0-to-1 scaling and the normalization have different properties, already illustrated above and further appreciated in the following test-particle approach, which makes them a great tool for comparing numerical and observational results. These two methods are easy to implement and come with barely any computational cost. Together with the interpolation approach, they may offer interesting applications for characterising the multi-dimensional VDF in other heliophysics and planetary physics contexts.

\section{Wave analysis and theoretical linear model}\label{sec:wave}

    \begin{figure}
        \centering
        \includegraphics[width=\textwidth]{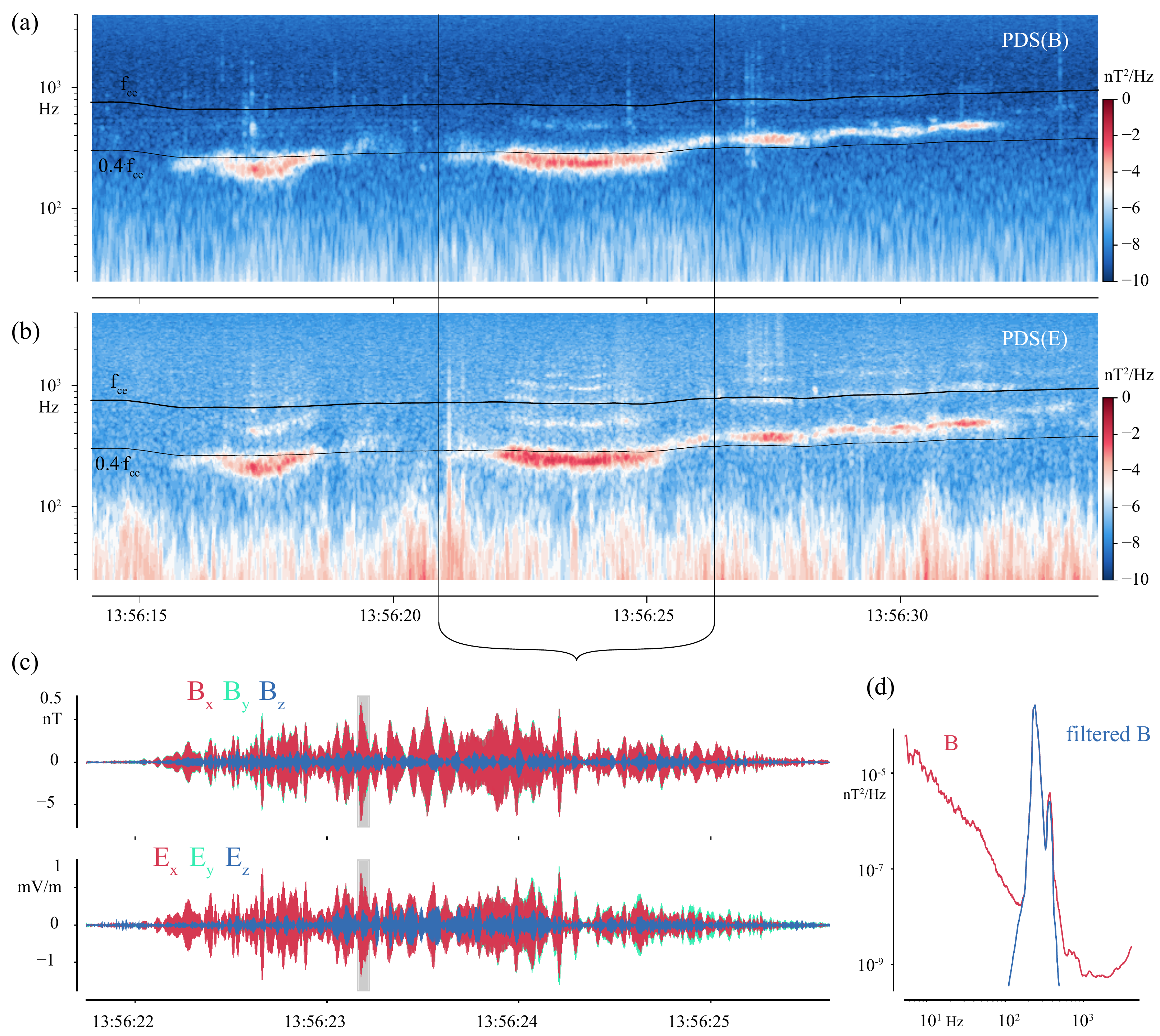}
        \caption{(a)-(b) B-field and E-field power density spectrograms. The electron gyrofrequency is indicated by the solid lines ($f_{ce}$ and $0.4 \ f_{ce}$). (c) Filtered E- and B-field wave forms. (d) Original B-field Power Density Spectra and its Butterworth filtered copy.}
        \label{fig:waveForm}
    \end{figure}

    In order to study wave-particle resonances in observational data, we have isolated one case displaying a clear wave activity within an otherwise ``quiet'' magnetosheath environment. Captured on the 8th of March 2019, the 3 minute-long interval starting at 13:56:10 displays contrasted, strong wave packets seen in the power density spectrograms of Figure \ref{fig:waveForm} (a-b). The central frequency of these packets is about 250 Hz, getting slightly higher on the second half of the observation. It is comprised between 0.3 and 0.5 electron gyrofrequency $f_{ce}$.

    We focus on the central event of constant frequency, selected in Figure \ref{fig:waveForm} (a-b). The fields components are band-pass filtered using a Butterworth filter. The result is displayed in Figure \ref{fig:waveForm} (d). The reference frame is aligned with the background magnetic field $\mathbf{B}_0$ averaged over the duration of the selected interval. We find that the parallel component of both fields is significantly smaller than the perpendicular components, however not null. In this frequency range, magnetic fluctuations up to 0.5 nT are observed around an average background magnetic field 25 nT strong (Figure \ref{fig:waveForm} (c)). The electric field fluctuations are seen sometimes surpassing 1 mV/m.

    \begin{figure}
        \centering
        \includegraphics[width=\textwidth]{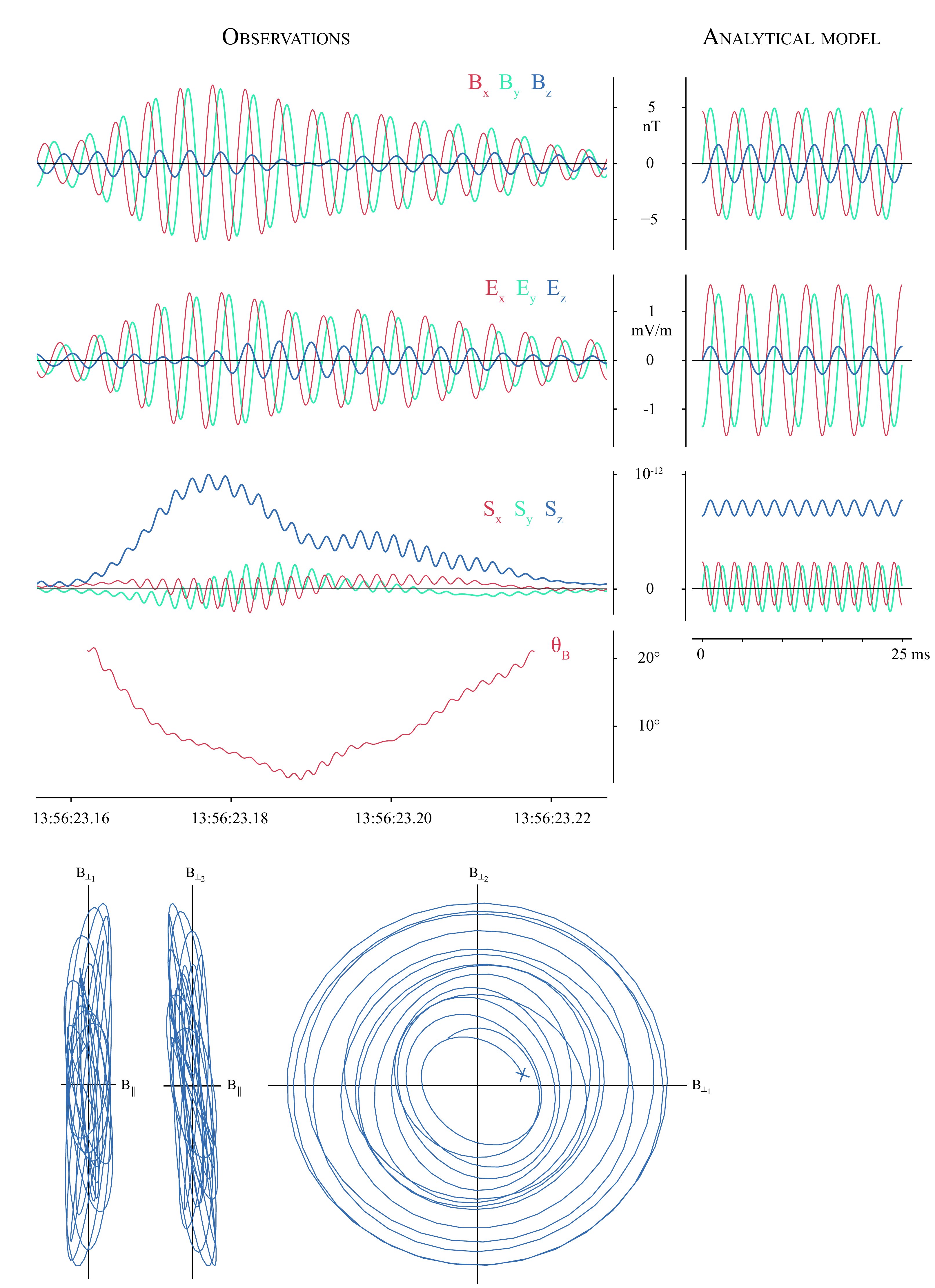}
        \caption{ One wave packet, indicated in Figure \ref{fig:waveForm} (c) with a grey background, its analytical model, at position (0,0,0), and its hodogram in the minimum variance analysis frame. The first data point of the series is indicated by a cross.}
        \label{fig:waveModel}
    \end{figure}

    We zoom-in once more to study a single wave packet, highlighted with a grey background in the time series of Figure \ref{fig:waveForm} (c) and expanded in Figure \ref{fig:waveModel}. The background magnetic field $\mathbf{B}_0$ is now calculated over this shorter period. The fields components exhibit remarkably clean sinusoids within the envelope defining the packet. The two perpendicular ($x$ and $y$) components of both fields are almost equal in magnitude and separated in time by a phase of $\pi/2$, corresponding to an almost perfect right-handed circular polarisation. The polarisation is also nicely seen in the hodogram of Figure \ref{fig:waveModel}, given in the local minimum variance reference frame (the frame in which the variance along an axis -- here the $z$-axis -- of the vector time series is the smallest, 65 times smaller than the two other variances, almost equal to each other).


    Whereas the four perpendicular components are defined by a similar envelope in the fields $\mathbf{E}$ and $\mathbf{B}$, the parallel (blue) component, also sinusoidal, follows a different time variation in $\mathbf{E}$ and $\mathbf{B}$, and has a slightly higher frequency compared to perpendicular components, in both fields (not shown). Given its right-handed circular polarisation and its non-zero parallel component, we identify the wave as a quasi-parallel whistler mode wave. Studying the same mode, \citeA{lacombe2014apj} and \citeA{stansby2016apj} report its observations in a range from 0.1 to 0.6 gyro-frequency, a range in which the present case falls perfectly. Because of the wave right-handed circular polarity and its frequency, electrons are the species of interest for this study.

    %


    We use the following dispersion relation of the whistler mode, derived from the Appleton-Hartree relation \cite{hsieh2017jgr} in the limit of $\omega \ll \omega_{pe}$, valid in our case with $\omega<10^{-2}\omega_{pe}$:

    \begin{equation}\label{eq:dispRelation}
    Y = \frac{X^2}{X^2+1}\cos(\theta)
    \end{equation}

    with the normalized spatial frequency $X=k\cdot d_e$, the normalized angular frequency $Y=\omega/\omega_{ce}$, $k$ the amplitude of the wave vector, $\omega$ the wave angular frequency, $d_e$ the electron inertial length, $\omega_{pe}$ the electron plasma frequency, $\omega_{ce}$ the electron (angular) gyrofrequency, and $\theta$ the angle between the background magnetic field and the wave vector.

    Because of apparent monochromatic nature of the observed waveforms, we can determine the wave vector angle $\theta$ with regard to the magnetic field using a minimum variance analysis over a sliding window as wide as the wave temporal period. In other words, we consider one pseudo-circle at a time, described by the wave vector, and find the orientation of the plane containing this circle. The result of this sliding minimum variance analysis is shown in the fourth row of Figure \ref{fig:waveModel} for the magnetic field. We find that the wave is propagating with angles between 0 and 20 degrees from the background magnetic field. Finally, to determine if this quasi-parallel wave is propagating along or against the background magnetic field, we calculated the components of the Poynting vector $\mathbf{E} \times \mathbf{B}$ , which is parallel to the wave vector (see for instance the comprehensive work of \citeA{stansby2016apj} on single-spacecraft estimation of whistler mode wave properties). The Poynting vector $\mathbf{S}$ (Figure \ref{fig:waveModel} third row) is mostly parallel to the background magnetic field, as evidenced by its largest and positive parallel component: the wave propagates in a mostly parallel direction. We now have the frequency and the wave vector direction of the wave, and an estimation of the wave vector amplitude. It was also verified that the probe velocity in the plasma frame (in which $\mathbf{u}=0$, see Section \ref{sec:partAnalysis}) is not significant with regard to the phase speed of the wave, and the observed wave frequency found in the spectrograms is not Doppler affected.

    Despite how clean the observed waves are, they are still limited to one point in space. If we are to study the possible resonant interactions between the particles and this precise type of wave, we need its temporal \emph{and} spatial description, or model. \citeA{hsieh2017jgr} developed an analytical expression for the components of a quasi-parallel whistler mode wave, based on the electric field linear system proposed by \citeA{stix1962} and obeying the Faraday's law. With the $k$-vector lying in the $(x,z)$-plane, the wave fields are given by the authors as

    \begin{equation}
        \begin{split}
            \mathbf{B}_w & = \mathbf{e}_x B_x^w \cos(\Psi) + \mathbf{e}_y B_y^w \sin(\Psi) - \mathbf{e}_z B_z^w \cos(\Psi) \\
            \mathbf{E}_w & = \mathbf{e}_x E_x^w \sin(\Psi) - \mathbf{e}_y E_y^w \cos(\Psi) + \mathbf{e}_z E_z^w \sin(\Psi) \\
            \Psi & = \omega t - k_x x - k_z z
        \end{split}
    \end{equation}



    Following \citeA{hsieh2017jgr} in the limit $\omega \ll \omega_{pe}$, we get the following polarizations

    \begin{equation}
        A_s = Y + \frac{Y^2-1}{\cos(\theta)-Y} \quad ; \ \
        A_p = \frac{\sin(\theta)\cos(\theta)}{\sin^2(\theta)-1+\cos(\theta)/Y}
    \end{equation}

    Finally, the wave fields components are expressed as


    \begin{equation}\label{eq:wave1}
        \begin{gathered}
        B_y^w =  A_s(1-A_p \tan(\theta)) B_x^w \quad   ; \ \ B_z^w = \tan(\theta) B_x^w , \\
        E_x^w = A_s  v_{p \parallel} B_x^w \quad ; \ \ E_y^w  = v_{p \parallel} B_x^w \quad  ; \ \ E_x^w = A_s A_p v_{p \parallel} B_x^w ,\\
        B_x^w =  \frac{B_w}{\sqrt{\cos^2(\Psi) + A_s^2(1-A_p \tan(\theta))^2 \sin^2(\Psi) + \tan^2(\theta)\cos^2(\Psi)}} ,
        \end{gathered}
    \end{equation}

    with the parallel phase speed $v_{p \parallel} = \omega/k_\parallel$ .

    The observed and modeled wave and plasma parameters for the packet shown in Figure \ref{fig:waveModel} are gathered in Table \ref{tab:param}. Using these parameters, we have estimated the wave vector amplitude using Equation \ref{eq:dispRelation}, also given in the same table. We can now obtain the fully analytical, temporal and spatial model of the wave, using Equation \ref{eq:wave1}. We note that no information about the electric field is fed to the analytical model, this field is thus completely constrained by the model.

    The result for a fixed point in space is given in the lower-right panels of Figure \ref{fig:waveForm} over a few wave periods. The analytical model results in a slightly higher $E_w / B_w$ ratio, but provides satisfactory wave forms and Poynting vector form and amplitude, additionally obeying Maxwell's equations. We can now use this model to investigate resonant wave-particle interactions.

    \begin{table}
        \centering
        \begin{tabular}{c|c|c}
              & Observed & modeled \\
            \hline
            $B_0$ & 25.4 nT & 25.4 nT \\
            $n_e$ & 16.3 cm$^{-3}$ & 16.3 cm$^{-3}$ \\
            $B_w$ & [0.2, 0.7] nT &  0.5 nT \\
            $E_w$ &  [0.5, 1.4] mV/m & 1.5 mV/m \\
            $\omega \ (\omega/\omega_{ce})$ &  1571 rad/s (0.35) & 1571 rad/s (0.35) \\
            $k$ ($k\cdot d_e$) & $\emptyset \ (\emptyset)$ & $5.9 \cdot 10^{-4}$ rad/m (0.77) \\
            $\theta$ & $[0^\circ, 20^\circ]$ & $20^\circ $ \\
            \hline
            $T_\parallel$ & 45 eV & 45 eV \\
            $T_\perp$ & 41 eV & 45 eV \\
            $\beta_e$ & 0.46 & $\emptyset$ \\
        \end{tabular}
        ~\\
        \caption{Wave parameters for the observation and the model, and additional plasma parameters.}
        \label{tab:param}
    \end{table}

\section{Resonant test-particles vs. observed VDF}\label{sec:testPart}

    We will now examine the potential effect of the modeled wave on the electron dynamics, and see whether or not signatures in the electron VDF can be found. Before considering simulating the situation with a self-consistent model (fields and particles feedback on each other according to Maxwell's equations), we explore the possibilities offered by a test-particle approach. The fields and waves are analytically described with constant frequency and wave-vector through time: there is no feedback from the particles on the fields. Thus, the presence of other particles, or the shape of the distribution, does not affect a single particle dynamics, which can be solved on its own. This allows for a very cheap, flexible first study of the resonant dynamics in the case of our simplified wave. A considerable advantage is that by doing so, we isolate the effects \emph{of} the wave \emph{on} the particles, and not the other way around, which is not directly possible in a self-consistent description.

    In the absence of collisions and gravity, and using the analytical wave discussed in the previous section, we know the force experienced by a particle at any time and any position. The particles dynamics are solved according to this force using the classical Boris scheme \cite{boris1970}, widely used by the Particle-In-Cell community because of its straightforward implementation and its great accuracy for this precise problem -- charged particles moving in electromagnetic fields. We initialise 200 million particles following a 3-dimensional isotropic non-drifting Maxwellian distribution, characterised by the observed electron (parallel) temperature of 45 eV. These particles are homogeneously distributed in physical space, in a 2-dimensional periodic box, extending over one projected wavelength (parallel for one axis, perpendicular for the other). Results are given for a simulation time of 25 ms, corresponding to a bit more than 6 wave periods and about 15 electron gyrations.

    \begin{figure}
        \includegraphics[width=\textwidth]{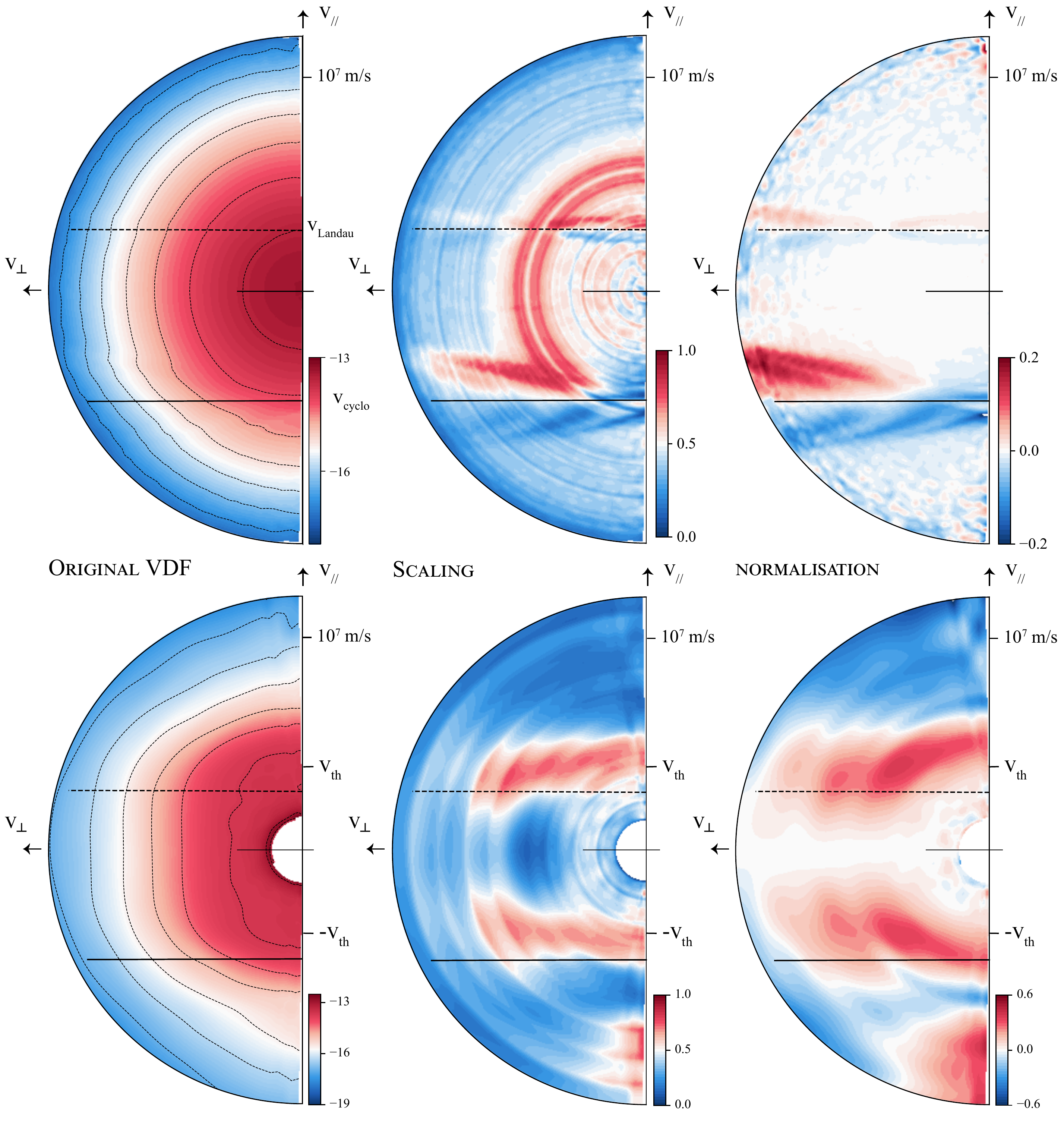}
        \caption{Test-particle results (top) compared to the observation (bottom), for an integration time of 60 ms. The Landau parallel resonant velocity is given by the horizontal dashed line and the cyclotron one by the solid lines. }
        \label{fig:comparison}
    \end{figure}

    The Landau resonance is obtained for particles with a parallel velocity close to the wave parallel phase speed: seen from these particles' perspective, the parallel component of the wave is almost static, and continuously accelerates them. This gives us a first resonant speed, $v_{Landau} = \omega/k_{\parallel}$. Because of their angle from the background magnetic field, the observed and modeled waves have a projected parallel component, which therefore enables this resonance. Particles with a parallel speed slightly slower than the parallel phase speed are being caught up by the wave and gain kinetic energy, whereas faster particles experience the opposite phenomenon. These resonant particles migrate in phase space, with a motion dependent on their initial phase space position. The wave we consider here has a dependence along the perpendicular direction: as a particle gyrates, its perpendicular position evolves, and so does the magnetic and electric wave components it experiences. A deeper, 3-dimensional description of these resonant motions, depending on $v_\parallel$ and $v_\perp$ is beyond the scope of this article, despite its great interest. These dynamics quickly result in a mixing of the particles in phase space around the parallel Landau speed. It is noteworthy that if one initialises the simulation with a flat velocity distribution (i.e. no density gradient in velocity space), no signature of this mixing can be found in velocity space. A density signature is indeed only visible if the resonant speed corresponds to high VDF gradients, which happens to be the case here. Very quickly, in just 6 wave periods, the resonant particles form an under/over-density centered on the Landau resonant speed, as seen in the scaled and normalized VDF of Figure \ref{fig:comparison} (the Landau resonant speed is displayed by the dashed horizontal line).

    The cyclotron resonance is only met in the presence of a circularly polarised wave. Particles with different parallel speeds experience a wave with a different frequency, a simple Doppler effect dependent on the particles velocity. For one particular parallel speed, given by $v_{Cyclo} = (\omega-\omega_{ce})/k_{\parallel}$, with $\omega_{ce}$ the electron gyrofrequency, the wave is seen with an angular frequency equal to the electron (angular) gyrofrequency: the particles gyrate synchronously with the rotation of the circularly polarised $\mathbf{B}_w$ and $\mathbf{E}_w$. Note that because $\omega<\omega_{ce}$, this resonant speed is negative (if $k_\parallel>0$): resonant particles are moving against the wave, which in turn significantly decreases the time during which they may interact with the wave. As discussed in the Landau case, such particles are continuously accelerated by the wave, and migrate in phase space, in an even more complex manner. The mixing of a high density gradient region again results in an under/over-density organised around the cyclotron speed given by the solid line in Figure \ref{fig:comparison}, with a sign opposite to the one of the Landau resonant speed. The over-density is found for lower parallel speeds but higher perpendicular speeds, corresponding to higher total speed for these particles. 

    The lower row of Figure \ref{fig:comparison} gives a comparison to the VDF observed during the same wave packet. We use only two FPI observations, corresponding to an integration time of 60 ms. There as well, two over-densities are found just above the two resonant speeds, at the same location as for the test-particles. The scaled view of the test-particle VDF exhibits two strong, perfectly circular artefacts, which were discussed in the first section. The normalized view, conserving the continuity along the radial dimension, does not display such structures. An additional feature is also found along the anti-parallel direction. We will see in the next section that this beam does not correlate with the presence of the wave. The thermal parallel speed is indicated in the observed distributions. The speed is the same as the central speed of the over-densities: the resonant mixing of particles indeed happens where the VDF velocity gradient $\partial f/\partial v_\parallel$ is high, which could lead to an efficient damping of the wave. We note however that the relevance of this thermal speed is limited, when considering the large departures of the observed VDF from a Maxwellian distribution.\\
    The over densities are observed around similar absolute parallel velocities, despite the absolute value of the resonant parallel velocities being clearly different: the over-densities tend to be fairly symmetric with regard to the $(v_{\parallel}=0)$-plane. Note that this fact holds for the plasma and wave parameters analysed here. If the position of the two over-densities are matching nicely between the test-particles and the observed VDF, strong discrepancies in the shape of these signatures exist and are discussed below.

\section{VDF time evolution}

\begin{figure}
        \centering
        \includegraphics[width=\textwidth]{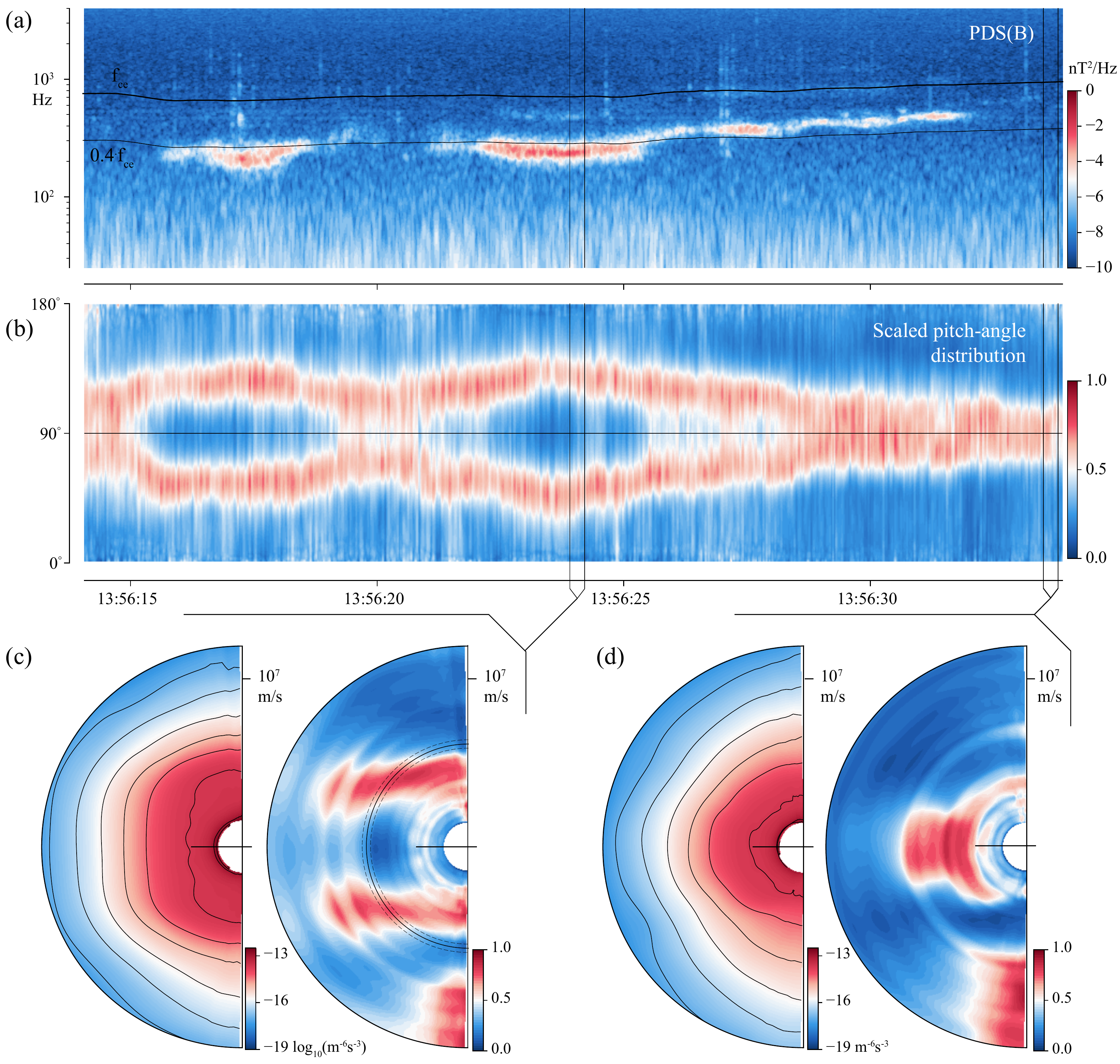}
        \caption{(a) B-field power density spectrogram. (b) scaled pitch-angle distribution for a speed range indicated in (c). (c)-(d) time-integrated VDFs and their scaled view, with the integration time indicated on the time series.}
        \label{fig:spectroVDF}
    \end{figure}

    To verify if this double branch signature indeed correlates with the presence of the wave over longer time scales, we show In Figure \ref{fig:spectroVDF} (c-d) two additional electron VDFs integrated over a longer time of ten FPI measurements, corresponding to an integration time of 300 ms. In both cases, a strahl-like (beam) component is found along the anti-parallel direction, with a constant perpendicular width. In the absence of wave activity (last VDFs), we find a clearly anisotropic distribution, with a strong equatorial signature in the scaled view.\\
    When analysed during the maximum power of the wave activity, the VDF displays the double-branch signature. It was checked that during the entire time interval, no significant (i.e. higher than the measurement noise) agyrotropy are to be found. Ions (not shown here) present a similarly anisotropic distribution, virtually static over the time interval.\\
    Figure \ref{fig:spectroVDF} (b) gives a view on the temporal evolution of the scaled VDF and its double-branch feature. For this representation, each 30 ms scaled VDF was averaged over a limited speed range between 5.8 and 6.3 $10^6$ m/s (indicated on the scaled view in panel (c)), and the result was plotted in the time $t$ and pitch-angle $\Theta$ dimensions in panel (b). Note that scaled pitch-angle distributions and classical pitch-angle distributions qualitatively converge as the energy range decreases. In this energy range, scaled values above 0.5 (red tones) form a striking feature centered on 90 degrees. When the waves are observed in the magnetic field spectrogram (panel (a)), the feature splits into two branches, corresponding to the double-branch pattern observed in velocity space. We note that such a pitch-angle distribution looks very similar to electron distributions in mirror modes, explored by \citeA{breuillard2018jgr}, where the envelope of the pitch-angle distribution is shown to correlate with a critical pitch-angle, under which electrons get trapped in one mirror structure. We note, however, that in our case, the signature is ordered by (constant) parallel velocities, and not constant pitch-angles when all the energy range is considered.

\section{Particle energisation}

\begin{figure}
        \includegraphics[width=\textwidth]{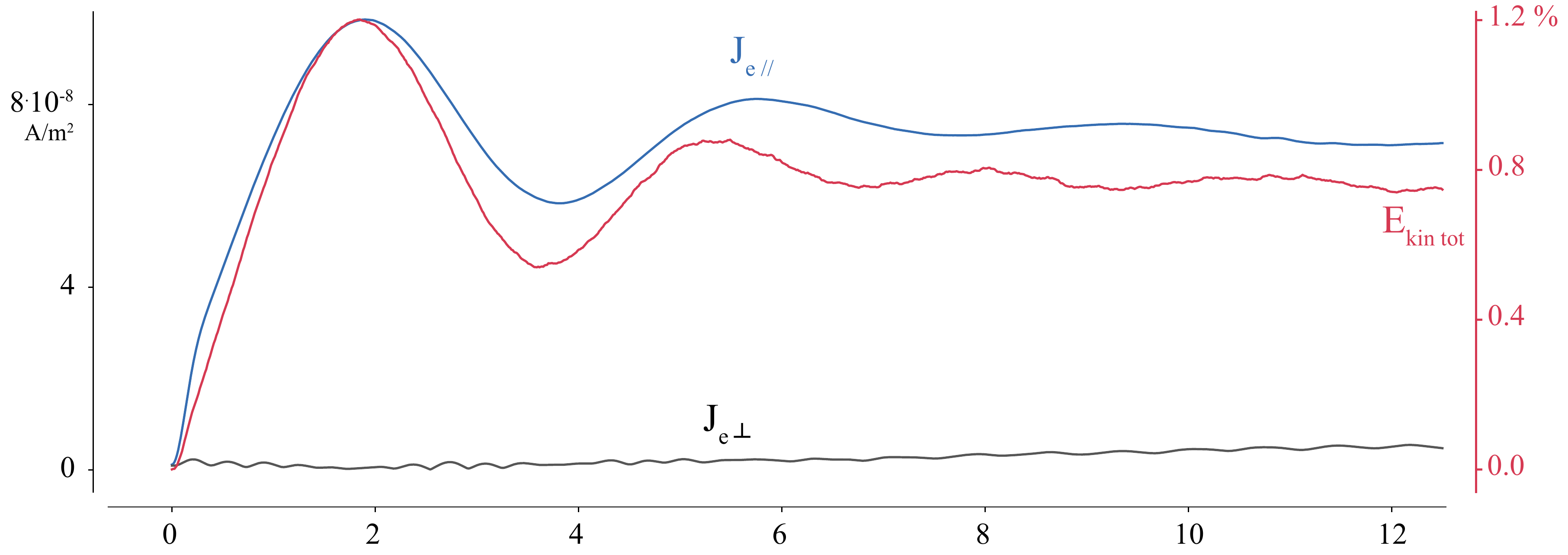}
        \caption{Test-particle total kinetic energy (relative to its value at t=0) and current density, parallel and perpendicular. Time is given in unit of wave period $T_{wave}$.}
        \label{fig:currentTestPart}
    \end{figure}

    We can follow the energisation of the test-particles either by summing up the total kinetic energy of all particles, or simply by taking the center of mass of their distribution (order 1 moment) and expressing it as a current density, with one parallel and one perpendicular component. The total kinetic energy and the two components of the current density are given in Figure \ref{fig:currentTestPart}. We find that with the physical parameters of our problem, the maximum energisation happens very fast, in about 2 wave periods, or 5 gyroperiods. Therefore, the wave packets are sufficiently long to exchange energy with the particles. After about 5 wave periods, the overall energy gained by the test-particles, as well as the parallel current density, stabilise to a non-zero value. The test-particle approximation does not hold anymore, as this current would necessarily alter the electric and magnetic fields, which would act to decrease it back to zero. For this reason, it is likely that this estimated current for the present resonant wave-particle interaction is an upper bound for the self-consistent interaction. Barely any perpendicular current is found, indicating that the VDF remains gyrotropic at all times. The fact that the total kinetic energy of all particles and their parallel current have almost the same time evolution illustrates that these resonant interactions mostly result in an increase of the bulk velocity and only some of the energy is transformed into ``temperature'', or change of the pressure tensor.

    A crucial point is indicated by this time evolution and verified in the time evolution of the simulated VDF (not shown) is that the resonant signatures are not periodic: they remain at the same position (in velocity space) with the same over/under densities through time. This is a necessary condition for it to be observed by an instrument with an integration time longer than the wave period.

\begin{figure}
        \includegraphics[width=\textwidth]{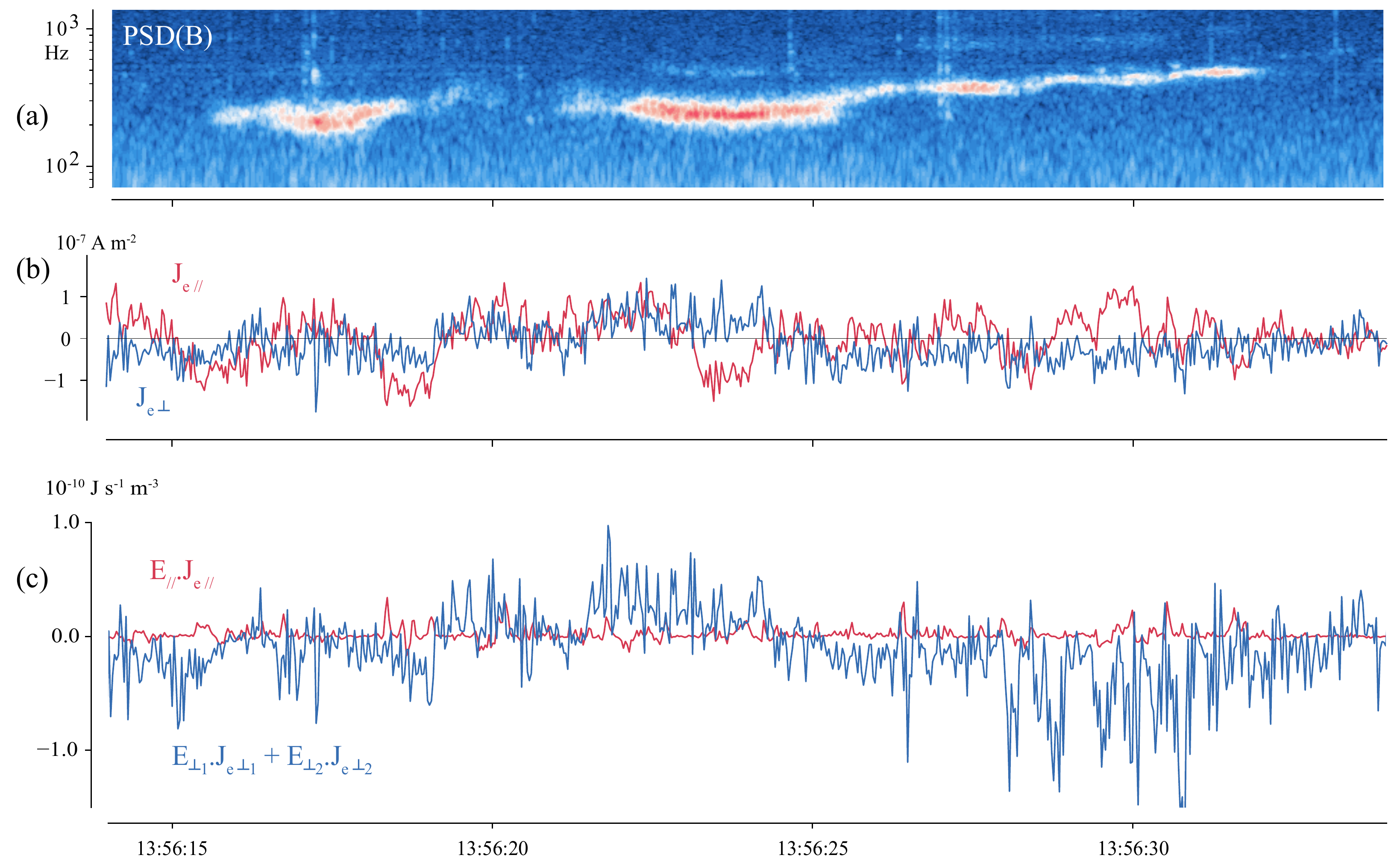}
        \caption{(a) Magnetic field spectrogram, (b) observed electron current and (c) dissipative term.}
        \label{fig:currentObs}
    \end{figure}

    In Figure \ref{fig:currentObs}, the observed electron current density (parallel and perpendicular) are displayed, together with the cropped spectrogram of the magnetic field. The parallel current is varying around 0, and no correlation with the wave activity can be found. Despite the strength and duration of the wave packets, the \emph{background} current density turns out to be significantly larger than the estimated (upper bound) current rising from the resonant particle signatures. Alternative moments were also calculated ignoring the core of the distribution (using different speed threshold), showing barely any changes in the electron current. In turn, the dissipative term $\mathbf{E} \cdot \mathbf{J}_e$, using an electric field averaged over the electron integration time, shows no correlation with the wave activity.

\section{Discussion \& Conclusions}

    We have first shown that VDF features originally hidden under high gradients of the background VDF can be highlighted using some scaling and normalization, without the use of a reference VDF $f_0$. We then isolated a strong, fairly narrow-band, quasi-parallel whistler mode wave, which could be analytically mimicked. We applied this analytical wave to a collection of test-particles to get a first sense of its potential effects on the electrons. We could map where and how particles resonate, and have found that the initial Maxwellian distribution is reshaped by Landau and cyclotron resonances. Test-particles and observed VDFs display two branches of higher density, around constant parallel velocities, in the scaled and normalized views of the VDF. The observed signatures were found to correlate nicely with the wave activity. Finally, this wave-particle interaction could not be detected in the observed current or the observed $\mathbf{E}.\mathbf{J}_e$ product, as both $\mathbf{E}$ and $\mathbf{J}_e$ (observed) present fluctuations of larger amplitude than the current and electric field used by or resulting from the model. Strong discrepancies between the simulation and the observations remain to be discussed.

    Firstly, the strongest discrepancy between the observed VDF and the simulated one is the strong symmetry of the observed signatures around the $(v_\parallel=0)$-plane, further illustrated by the time evolution of the scaled pitch-angle distribution in Figure \ref{fig:spectroVDF}. As such, it is difficult to prove that the Landau resonance, providing an already faint signature in the perfect test-particle set-up, may cause a resonant branch as strong as the observation, and so symmetric to the cyclotron branch. It actually \emph{appears} as if a mirrored cyclotron branch shows up systematically with the wave activity, despite our wave analysis only bringing out one single mode, propagating in one well defined direction. Such symmetric signatures have been displayed in the works of \citeA{min2014jgr} (in the magnetosphere) and \citeA{chen2019nc} (in a nominal, or fully developed turbulence case), in which the authors did not isolate one single mode, and interpret the symmetric signatures as caused by waves propagating in both directions, parallel and anti-parallel. This symmetry remains the foremost open question of our study.

    Secondly, the test-particle model cannot account for the transformation from the already strongly anisotropic VDF out of the wave activity (shown in panel (d) of Figure \ref{fig:spectroVDF}) to the 2-branch signature. The disturbance of the VDF caused by resonances in the test-particle case is a local phenomenon in velocity space, it certainly cannot reshape macroscopically the distribution, make a large amount of the electron population migrate over large velocities. Were we able to easily initialise the particles according to the observed VDF outside the wave activity, the strong equatorial structure would remain there, with additional resonant signature expressed on the sides. While the test-particle simulation enlighten us on the micro-physics of the interaction during a very brief snapshot of the observations (when an intense wave activity is observed), the macroscopic configuration of the fields (magnetic gradients, mirrors, etc) should be considered to fully understand the time evolution of the electron VDF.

    The limitations of the test-particle simulation are numerous, if one is to thoroughly compare its results and the observations. Instead, we suggest that this approach only points at the very first step of the full self-consistent wave-particle interaction. It shows us where and how in velocity space resonances should occur for a given a monochromatic wave, and gives a first hint on their phase density signatures. Most importantly, it also gives us a good sense of how fast these resonances can produce signatures in the VDF, and their efficiency. But it obviously cannot go further in the physics of the interaction, both an advantage and a drawback.\\
    This numerical approach is limited to the case of a purely monochromatic wave with a constant amplitude and a constant normal angle, whereas the observed wave packets exhibit some spectral breadth, a parallel component at a slightly higher central frequency, and additionally a propagation direction constantly evolving. Because of the periodic boundaries necessary for the simple test-particle approach, such a wave cannot be easily modeled. We refer to the work of \citeA{tao2012grl}, in which the authors specifically highlighted the effect of the amplitude modulation of chorus emissions in the magnetosphere, on particle acceleration. They, however, do not reconstruct distribution functions, and can therefore do without the periodicity constraint.\\

    The feedback of the particles on the wave is a different problem entirely, in terms of numerics at least. Thus the comparison should not be over-interpreted. We have shown that the observed wave is expected to have clear signatures on the electron distribution function, signatures matching well in position and somewhat in shape the observed signatures.\\
    ~\\

    The visualisations (scaling and normalization) developed for this study are a key element for a better, deeper characterisation of the particles distributions. It is in the numerical aspect of our work that these views reveal best the effect on the distribution, getting us rid of the need for an additional information (e.g. distributions at other times). They show great flexibility and versatility, which may be useful for other applications on observations and simulation data, and mostly their comparisons. \\

    Such an exploration of wave-particle resonances in 3 dimensions, and its direct comparison to observations, is -- to the best of our knowledge -- novel. It familiarises us with resonant dynamics and signatures, at a low cost and without the intrinsic complexity a self-consistent model adds to physical interpretations. This approach may present real promises for a more systematic recognition of resonant signatures in a more complex data, such as those of fully developed turbulence in the solar wind or the magnetosheath

\acknowledgments
E. Behar is funded through DIM-ACAV post-doctoral fellowship and from the European Unions Horizon 2020 research and innovation program under grant agreement No 776262 (AIDA, www.aida-space.eu). The authors also wish to thank Olivier Le Contel and G\'erard Belmont for valuable discussions that have broadened the context of this study.

\newpage
\appendix

\section{Binning versus interpolating}\label{app:binningInterpolation}

    \begin{figure}
        \centering
        \includegraphics[width=\textwidth]{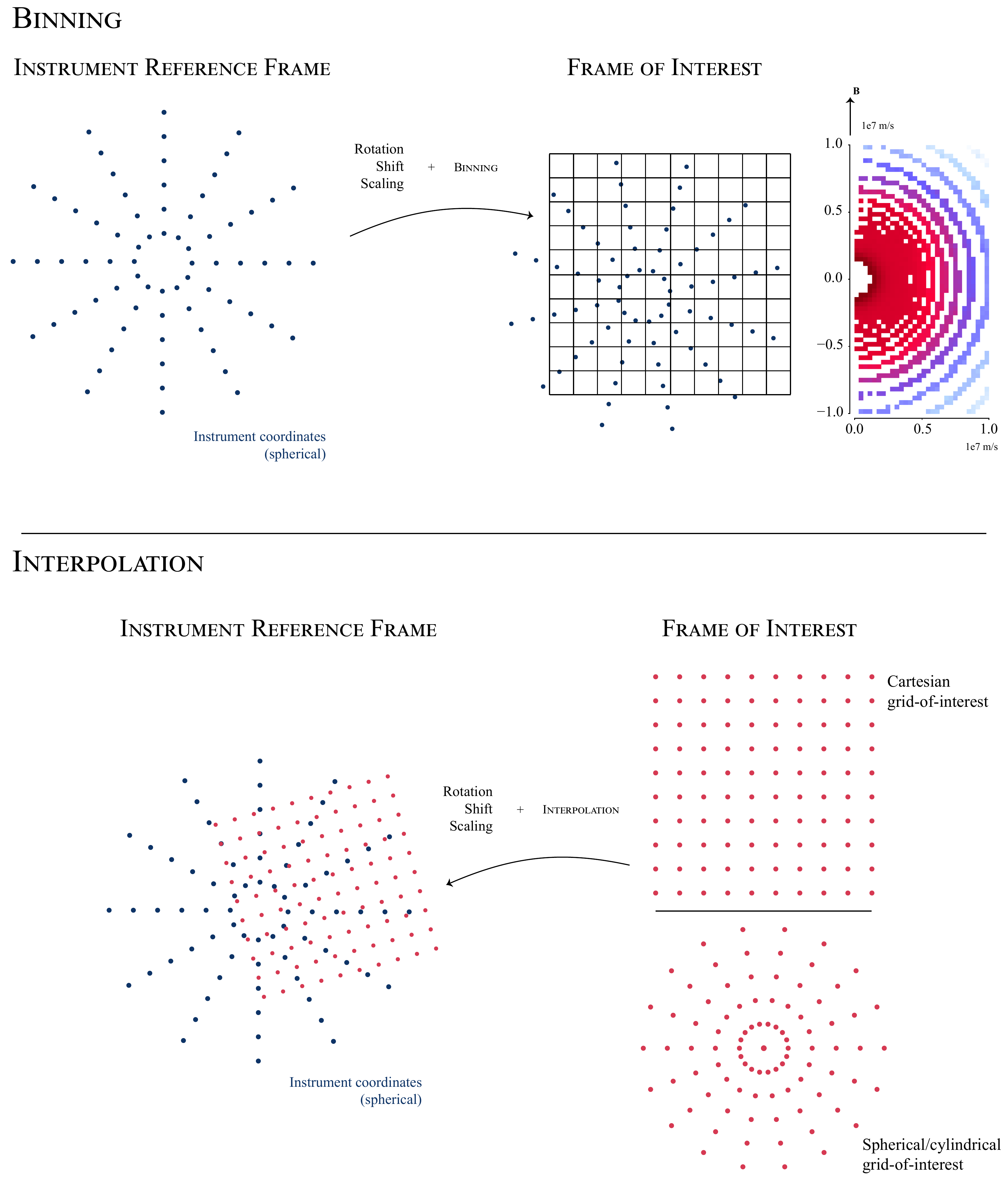}
        \caption{Illustration of a binning and an interpolation.}
        \label{fig:interpolationSchem}
    \end{figure}

    In various publication, when handling multidimensional Velocity Distribution Functions (VDF), authors choose the somewhat more intuitive of rotating the data from the instrument frame to the frame of interest, to then \emph{bin} the data within a defined grid, as illustrated in Figure \ref{fig:interpolationSchem} in the first row. During this process, one measurement point ends up in one single cell of the grid-of-interest, and one cell of the grid-of-interest may receive zero one, or many measurements. Empty cells will appear on the representation, producing visual artifacts, and it is necessary to average the binned data over at least one dimension in order to \emph{fill} as much as possible the grid-of-interest. We give such an example in a cylindrical representation of the VDF, right-most plot of the first row of Figure \ref{fig:interpolationSchem}, which shows about thirty electron distributions binned in a 2-dimensional grid defined, in the classical $(v_\parallel, v_\perp)$-plane. The binned data show strong artifact along the energy dimension, as the instrument energy levels are log-distributed: the higher the energy, or speed, the greater the energy steps. Therefore one regular Cartesian grid cannot be suitable for the entire energy range of the instrument: its resolution will be too coarse at low energies -- where weighting of the binned data has to be taken into account properly -- and too fine for high energies, leaving empty most of the grid-of-interest. The binning approach can only be used over a restricted range of energies, and is anyway not suitable for 3-dimensional analyses. \\

    The interpolation approach is illustrated in the second row of Figure \ref{fig:interpolationSchem}. First, we define a 3-dimensional grid-of-interest -- an array/set of coordinates -- within the reference frame of our choice, with arbitrary extent and resolution. We apply the opposite rotation, shift, and scaling to a copy of the grid-of-interest, from the reference frame to the instrument frame, as illustrated in Figure \ref{fig:interpolationSchem}. The values of the data are then interpolated at each node of the transformed grid (see next section): the data are continuously evaluated over the entire grid, leaving no room for artifacts, and avoiding additional weighing operation. All the results in the article use a spherical grid-of-interest, allowing at no additional cost the scaling and normalization of the VDF, and their straightforward representation in the $(v_\parallel, v_\perp)$-plane.

    But interpolation with an order higher than 2 presents complications, and this approach is not universal and cannot be applied blindly.

\section{Interpolating velocity space distributions}\label{app:interpolation}

	We have tested three different interpolation schemes, namely nearest-neighbour (order 0, only one point of the measured VDF is used), trilinear (order 1, 8 data points are used), and tricubic (order 2, 64 data points are used). The two first schemes are taken from the main scientific python library, SciPy, while the third was implemented by \cite{lekien2005nme}, with one wrapper made available by the authors for a usage with Python.

	\begin{figure}
        \centering
        \includegraphics[width=\textwidth]{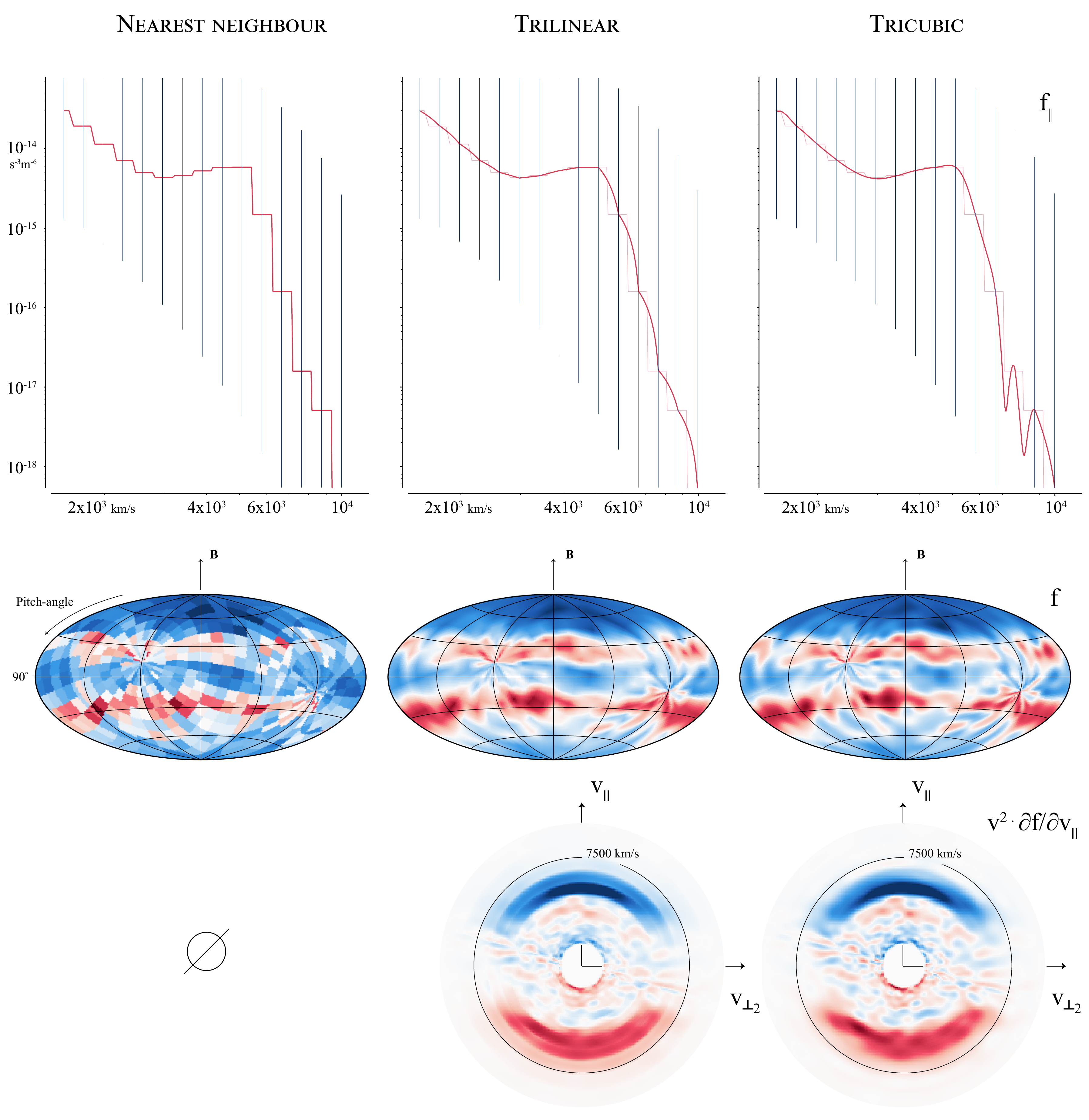}
        \caption{Interpolation.}
        \label{fig:interpolation}
    \end{figure}

	We now have a 3-dimensional array of interpolated VDF values, with two different sets of coordinates, one in the instrument frame (used for the interpolation), one in the reference frame of the study (the originally defined Grid-of-Interest). This array can be either analysed and visualised on its own, or added up to other arrays in order to average the data over a longer duration, all within the same frame. For instance, Figure \ref{fig:interpolation} presents results using only one single electron VDF, which is the most challenging test for the methods, because of low detection rates for higher energies. 

	This method has one invaluable interest: any resolution can be used without any of the risks inherent to a binning process, illustrated in Figure \ref{fig:interpolationSchem}. Another great advantage when using a Cartesian grid-of-interest is that velocity derivatives of the data can be easily and straightforwardly computed. The advantage of using a spherical or cylindrical grid-of-interest was already illustrated. The method has one drawback, namely the computational cost. The interpolation in itself is more computationally demanding than a simple binning, and for most purposes we will have many more elements in the grid-of-interest than in the instrument grid, resulting in as many more calculations to compute.\\

	In Figure \ref{fig:interpolation}, a single electron distribution is used to illustrate the differences between the three interpolation schemes, using three different visualisations. In the first row, we give the profile (1-dimensional) of the interpolated values along the parallel direction. For this representation and in order to increase slightly the statistics, all values within an angular distance from the parallel axis ($+\vv{B}$) are selected and averaged, using a conic selection $\pi/6$-wide. For this purpose, a spherical grid-of-interest is a better choice, allowing one to simply average the interpolated data over one angular dimension, avoiding a binning process. 

	The energy/speed levels of the instrument are also indicated as vertical lines, regularly spaced in this logarithmic representation. For this analysis, we only rotated the frame, without a shift, so the energy levels of the instruments remain centered on the origin of the Grid-of-Interest, making the results more readable.\\

	Another representation is proposed on the second row of Figure \ref{fig:interpolation}, in which we selected a specific speed range, averaged the data over this range, resulting in the given angular maps (2-dimensional). Here as well, using a spherical Grid-of-Interest makes this selection straightforward, selecting only one radial range in the array, again avoiding a binning process. The parallel axis is vertical and intersects these maps at their poles, and the equator correspond to a pitch angle of 90 degrees. \\

	In the parallel 1-dimensional profiles previously described, the nearest-neighbour interpolation results in steps centered around the instrument energy levels, as expected from such a scheme. This curve is reported in the two other VDF profiles for the trilinear and tricubic cases, in order to verify that all three interpolations indeed meet at the instrument speed levels, an important convergence test. These steps in the nearest-neighbour case directly correspond to the ``pixels'' found in the left-most angular map. This is the way particle data are often displayed, implying that within an instrument ``pixel'', the observed flux is constant. This assumption is also widely used when integrating the plasma moments, which simply corresponds to a Riemann sum of the area under a function. The instrument poles are misaligned with the magnetic field direction, and can be seen close to the equator, 180 degrees apart, with triangular pixels meeting in one point.

	The trilinear interpolation provides a smoother, more continuous profile. Thinking in one dimension, between two measurement points, interpolated values will follow a linear relationship. In this logarithmic representation, these linear segments show up as arc segments in-between each instrument energy/speed level. This is obviously an unwanted visual artifact, generated by the choice of the representation. Just as the 1-dimensional profile, the angular map shows a smoother result than in the nearest-neighbour case. In the same way that the nearest-neighbour can be linked to a Riemann sum, this linear interpolation converge to a trapezoidal rule in terms of VDF integral.

	Finally, the tricubic interpolation provides the smoothest and most continuous curve for speeds up to 7000 km/s, for which the VDF is relatively high. At higher energies, strong artifacts are found, with oscillations between the last instrument speed levels. Two phenomena can account for this behaviour. The first could be the Runge's phenomenon, namely oscillations at the edges of the interpolation interval when using polynomials of high degree (higher than two). We note however that we only use a third degree polynomial interpolation, which should limit this phenomenon, and these oscillations are only visually found for low fluxes, high speeds. The second phenomenon occurs when the VDF contains zero values. But most of all, this interpolation scheme, as derived and implemented by \citeA{lekien2005nme}, can result in negative VDF values. This nonphysical result should be monitored, so these negative values remain small and insignificant for the analysis we want to perform.\\

	\emph{Remark}: these interpolations and their unavoidable artifacts and drawbacks are not well suited for all purposes. For instance, the tricubic interpolation is valid only for strong signals, high VDF values, and should anyway be used with great caution, so these artifacts are not interpreted as physical features.\\

\begin{figure}
        \centering
        \includegraphics[width=\textwidth]{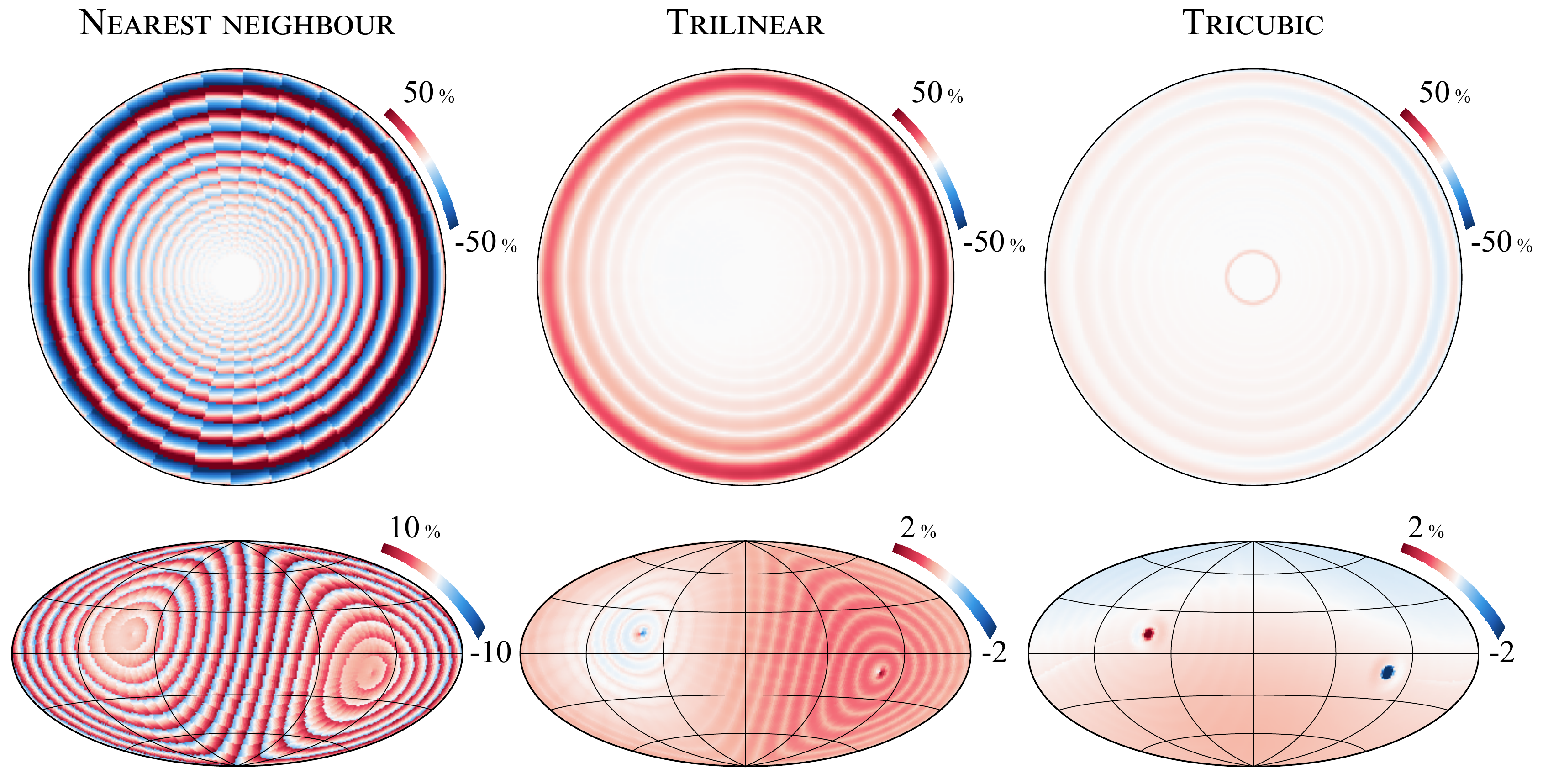}
        \caption{Error interpolation.}
        \label{fig:errorInterpolation}
    \end{figure}

    As a first test, we have already verified that the three schemes converge at the instrument speed levels. A second important test is to make sure that the three interpolations do not show artifacts at the instrument poles. Indeed, the trilinear and tricubic schemes we use assume that the data are defined over a grid with cuboid elements: evenly spaced along each dimension, with possibly different spacings along each dimension. The poles of the instrument spherical grid should therefore present errors, as the array elements there strongly depart from cuboids. To test the overall error from each scheme, we have defined artificial, ideal instrument measurements, namely a drifting\footnote{The drift is important for the significance of the test, as it misaligns the center of the artificial distribution, and the center of the instrument grid, testing the errors over all three dimensions.} Maxwellian distribution. This way, we also know the real value of the distribution at the nodes of the rotated, shifted, scaled grid-of-interest, and can compare these analytical values to the interpolated ones. It allows us to creates error maps, presented in Figure \ref{fig:errorInterpolation}. The first row shows one chosen Cartesian cut through the error distribution, while the second row gives one angular cut, or angular map. The nearest-neighbour interpolation results in the largest errors, just as expected from the steps seen in the 1-dimensional profiles of Figure \ref{fig:interpolation}: the interpolation is alternately much smaller and much larger than the real analytical value of the distribution. These errors can largely surpass the 50 percent level, because of the steepness of the VDF. Since the Maxwellian distribution is convex for high energies/speed, the error given by the linear interpolation is positive-only for high speed values, as seen in the first row, middle panel. At these higher speeds, the error can reach up to 50 percent, whereas at lower speed values -- of greater interest for us -- the error becomes arbitrarily low. The tricubic interpolation shows the smallest error in this plane with mostly positive values. We note a slightly stronger error-ring at the lowest speeds covered by the instrument, which we link to the Runge's phenomenon. At the high VDF values usually observed at these energies, this error may become significant for some purposes, though it has not been found to be the case for our analyses.

    The instrument poles are not intersected by this plane, but we expect to find the greater error there. They however show up in any angular map, as seen in the second row of Figure \ref{fig:errorInterpolation}. The poles still have the same position as previously, and can be easily spotted in each interpolation error map. The worst qualitative result is obtained for the tricubic interpolation, with relatively strong errors localised at the instrument poles, though the absolute error remains low (a few percents) and indeed extremely localised. We conclude that for our purpose, the three schemes are sound, resulting in acceptable errors where the observed VDF values are high, i.e. for speeds lower than $10^4$ m/s.

\begin{figure}
        \centering
        \includegraphics[width=\textwidth]{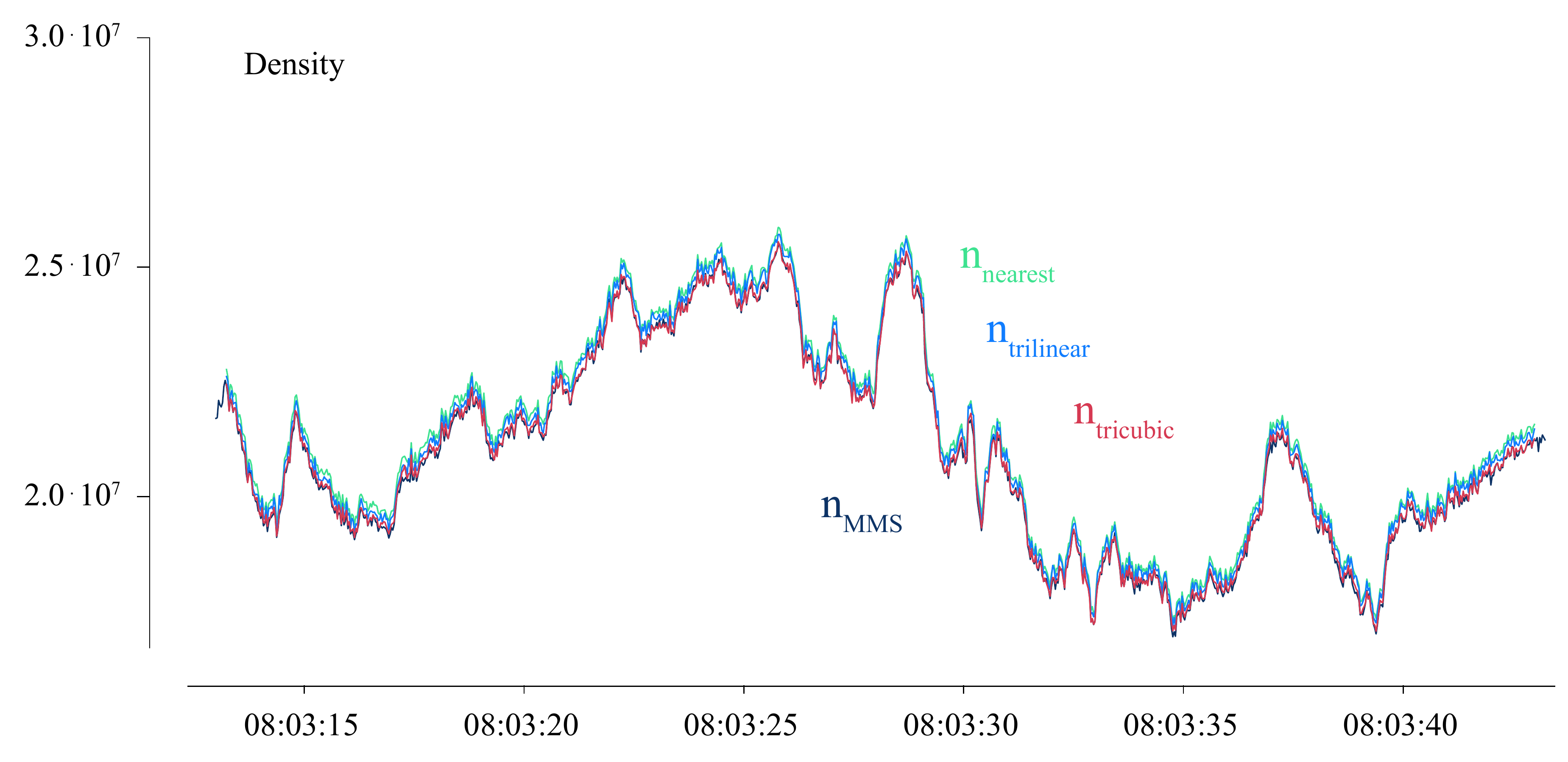}
        \caption{Moments tests.}
        \label{fig:momentsTest}
    \end{figure}

	As a third test, we have verified that the electron density, resulting from integrating the interpolated values using a simple Riemann integral, calculated for the three interpolations, matches the density provided by the instrument team. Their densities are also the plasma moment of order 0, but directly integrated in the instrument spherical coordinate system. The result is shown in the first row of Figure \ref{fig:momentsTest}, and one can find that the four curves are barely distinguishable.

\bibliography{libRes}

%
%
%
%
%

\end{document}